\documentclass[12pt]{iopart}
 \pdfoutput=1 
\usepackage{graphicx}
\usepackage{amssymb}
\usepackage{color}
\usepackage{cite}   
\DeclareGraphicsExtensions{.pdf,.png,.jpg} 

\begin{document}
\newcommand{\wq}[1]{\textcolor{blue}{#1}}

\title{Local Einstein relation for fractals}

\author{L.~Padilla,  J.~L.~Iguain}

\address{Instituto de Investigaciones F\'{\i}sicas de Mar del Plata (IFIMAR) 
and\\
Departamento de F\'{\i}sica FCEyN,
Universidad Nacional de Mar del Plata,\\
De\'an Funes 3350, 7600 Mar del Plata, Argentina}

\ead{iguain@mdp.edu.ar}

\begin{abstract}
We study single random walks and the  electrical resistance
for fractals obtained as the limit of a sequence of periodic structures. 
In the long-scale regime,  power laws describe
  both the mean-square  displacement of a random walk as a function of time and the electrical resistance as a function of length. We show that
  the corresponding power-law exponents satisfy the Einstein relation.
For shorter scales, where these exponents depend on length, we find how
the Einstein relation can be generalized to hold locally. All these
  findings were analytically derived and confirmed by numerical
  simulations.
\end{abstract}

\maketitle

\section{Introduction}\label{sec:intro}

Fractals are characterized by quantities that exhibit power-law behaviour 
in space or time.
More precisely, as scale invariance occurs for
integer powers of a characteristic length, pure power laws are 
modulated by logarithmic periodic functions, that
describe the departures from the main trend at intermediate scales. 
These modulations have been the object of recent interest
and considerable effort has been devoted toward understanding
the relation between log-periodicity and discrete-scale invariance~\cite{Grabner1997, acedo2000, bab2008EPL, bab2008JCP, 
maltz2008, weber2010, Padilla2009, Padilla2010, Padilla2011, havlin2000, bernhard2004, 
Padilla2012,refId0}.

For a given fractal and  some related observables, which show
 (modulated) power-law behaviours, 
a problem of interest is to determine
whether or not the exponents associated with these quantities are independent.
Sometimes we can expect a relation as a consequence of underlying
physical laws. This is, for example, the case of the mass $m$,
the electric resistance $R$ and the mean-square-displacement
(MSD) $\Delta r^2$ for a single random walker. On a fractal,
the first two
grow with length $l$ as $m(l)\sim l^{d_f}$
and $R(l)\sim l^\zeta$, while the last one grows with time $t$ as
$\Delta r^2(t)\sim t^{2/d_w}$. The exponents $d_f$, $\zeta$ and $d_w$
are known as the {\it fractal}, {\it resistance} and {\it walk} exponents,
respectively, and these power-law behaviours hold for 
scales large enough to ensure self-similarity.
In an $d$-dimensional euclidean space, the diffusion coefficient $D$ and
conductivity $\sigma$ are related by the Einstein equation
\cite{havlin1996} 
\begin{equation}
  \sigma=\frac{e^2\rho}{k_BT} D.
  \label{Eins_0}
\end{equation}
Here, $D=\lim_{t\to\infty}\Delta r^2(t)/2t$, $\rho$ and $e$ are the density
and charge of mobile particles, $T$ is the temperature and $k_B$ is the
Boltzmann constant. Equation (\ref{Eins_0})  is one of the forms of the
fluctuation-dissipation theorem, and can be used together with simple
scaling heuristic arguments, to argue that  the fractal, walk, and
resistance exponents satisfy the {\it Einstein relation}~\cite{havlin1996}
\begin{equation}
  d_f=d_w-\zeta,
  \label{Eins_1}
\end{equation}
This property has been shown to hold 
asymptotically for some finitely ramified 
fractals~\cite{Given_1983,Franz_2001}; which has been 
used to analyze the periodicity of the
oscillations in dynamic observables, in the first attempts to 
understand log-periodic modulation\cite{Maltz_2008}.
Einstein relation was also investigated for random walks on weighted
graphs~\cite{Telcs2006}, and,  more recently, for karst networks
structures~\cite{Hendrick_2016}.

A deterministic fractal can be obtained as the limit of 
a sequence of periodic structures. In this procedure, the period increases
at every step as $L^n$ ($n=0,1,2,...$),
where $L$ is a basic characteristic length scale. Self-similarity is 
manifested in power-law behaviours, which occur for long
enough scales. However, this does not always hold  for
shorter lengths. Thus, the local slopes of the observables as a function of time or
length, in log-log scales, are variable quantities, which approach 
constant values only asymptotically. 

In this work we argue that the local
fractal, walk, and resistance exponents are related through an equation 
that generalizes (\ref{Eins_1}).  This generalization is obtained 
analytically, following the steady-state method for the calculation 
of the effective diffusion coefficients for periodic
substrates~\cite{Aldao_1996}. To further strengthen our findings
we perform numerical simulations for two models of fractals; which
confirm the theoretical predictions. 

The paper is organized as follows. In Sec.~\ref{sec:perio} we relate
the diffusion coefficient and the unit cell resistance  for a periodic structure. In Sec.~\ref{sec:fractal} we derive the Einstein relation for self-similar systems. In
Sec.~\ref{sec:local} we generalize this relation for scale-dependent exponents.
In Sec.~\ref{sec:numeric} we confirm the generalized relation by numerical simulations
performed on models of asymptotic self-similar substrates. Finally, 
we give our conclusions in Sec.~\ref{sec:conclu}.

\section{Periodic systems}\label{sec:perio}

\begin{figure}[!ht]
	\begin{center}
\includegraphics[scale=.6,clip]{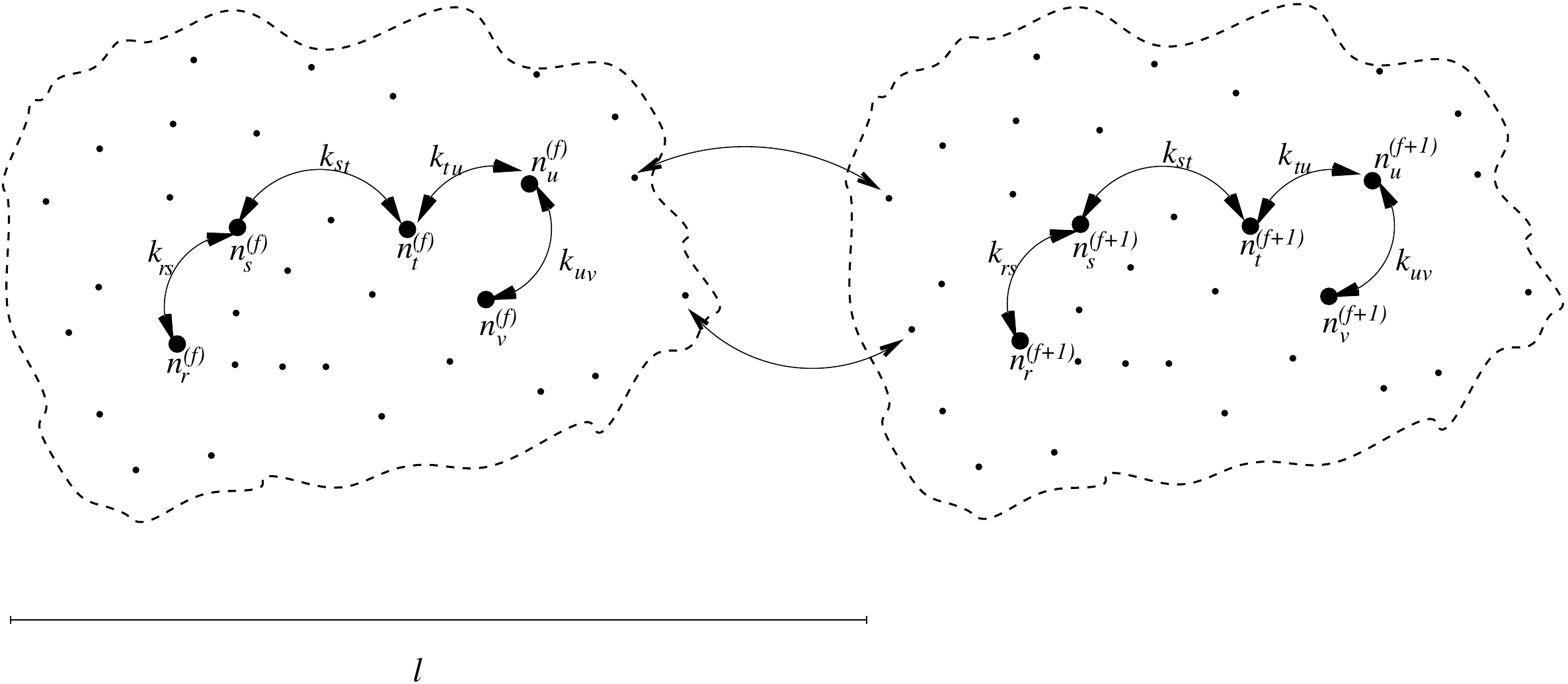}
 \end{center}
	\caption{Two nearest-neighbor cells $f$ and $f+1$, for a
        periodic substrate with linear size period $l$.  The
        points represent
        sites, which can be occupied by mobile particles.
        The arrows represent hopping rates between pairs of sites. For clarity,
        only a few sites and hopping rates were highlighted.
        $n_r^{(f)}$ corresponds to the number of particles in the
        internal site $r$ of cell $f$
	}
\label{cells}
\end{figure}

In this section we address the problem of the diffusion coefficient for
a periodic substrate. We follows the steady-state method developed in
reference \cite{Aldao_1996}. 
We start by introducing the periodic substrate  with  unit cell 
of linear dimension $l$, schematized in
figure \ref{cells}, where the points represent sites, and
the arrows 
represent hopping rates. On this structure, a mobile
particle can jump between connected sites according to the hopping rates $k's$ (for the sake of clarity only a few sites and arrows were highlighted).
We focus on a steady-state of non-interacting particles flowing  
with a constant current density $j$. 

As shown in \cite{Aldao_1996}, this steady-state consists of a set of microscopic currents distributed with the same
periodicity as the substrate. In figure \ref{cells} two nearest-neighbor
(NN) unit
cells are depicted schematically where, for example, $n_s^{(f)}$ represents
the number of particles in site $r$ (internal index) of cell $f$. Because of the  mentioned
periodicity, we  get that for given pair of 
connected sites with internal indices $s$ and $t$,

\begin{equation}
    i_{r s}^{(f)}= i_{r s}^{(f+1)},
\end{equation}
where $i_{r s}^{(f)}$ is the current from  site $s$ to site $r$ in cell
$f$. 
In addition, as hopping rates do not depend on the cell either  but
only on the internal indices, the last equation can be rewritten as

\begin{equation}
  k_{sr}( n_s^{(f)}-n_r^{(f)}) = k_{sr} (n_s^{(f+1)}-n_r^{(f+1)}),
\end{equation}
or

\begin{equation}
    n_s^{(f+1)}-n_s^{(f)} = n_r^{(f+1)}-n_r^{(f)}.
\end{equation}

Therefore, in the steady-state, the difference in the 
occupation number for a given site and the equivalent site in a NN cell is the same for all sites.

The relation of the steady-state problem with the diffusion coefficient
$D$ is provided  by Fick's law 
\begin{equation}
    j=- D\frac{\Delta n}{l^2},
  \label{ec_difu}
\end{equation}
which is valid for distances larger than $l$. 
Here $\Delta n$ corresponds to the particle number difference
for NN cells. Note  that  $D$ 
also determines the mean-square displacement $\Delta^2 x$ of a single 
random walker on the same structure, which behaves as
\begin{equation}
    \Delta^2 x(t) = 2 D t;
    \label{dx2}
\end{equation}
 for time long enough for $\Delta x \gg l$.

\begin{figure}[!ht]
	\begin{center}
    \includegraphics[scale=.6, clip]{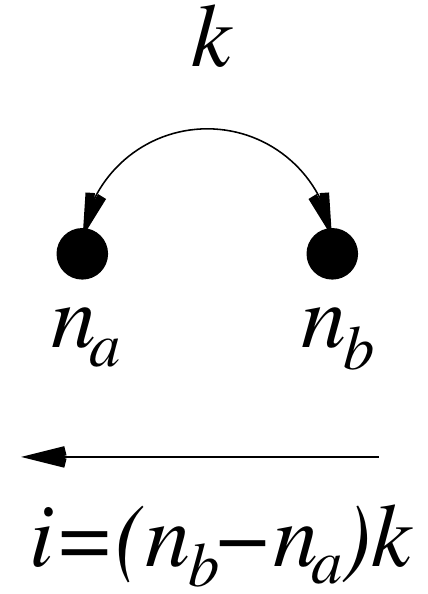}
		\hspace*{3cm}	
	 \includegraphics[scale=.6,clip]{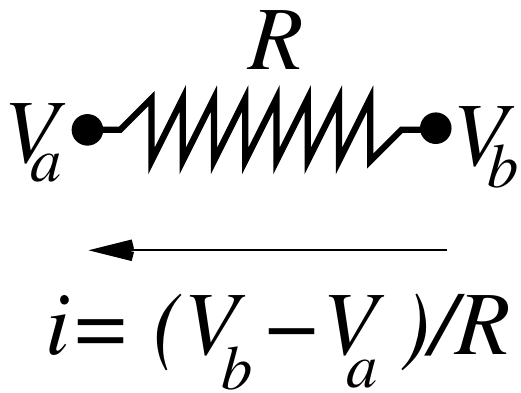}	  
 \end{center}
    \caption{Schematics of the equivalence between Fick's law (left) and 
    Ohm's law (right). In the mapping particles have unitary charge,
    while the other quantities are related as $V=n$, and $R=1/k$.
	}
\label{esque_equi}
\end{figure}
Transforming the steady-state problem into an equivalent electrical problem is straightforward.
Indeed, for particles of unitary electric charge, a mapping between
Fick's law and Ohm's law  
results by identifying particle number
with electrostatic potential ($V_a= n_a$) and hopping rate 
with conductance ($k= 1/R$). In figure \ref{esque_equi} we represent
this mapping for every pair of connected sites. 
Following this analogy, we see that in the electric problem, the potential
difference for a pair of equivalent sites in NN cells takes
the constant value

\begin{equation}
    \Delta V= n_r^{(i+1)}-n_r^{(i)},
\end{equation}
and that the difference between particle populations 

\begin{equation}
    \Delta n= \sum_{r=1}^M (n_r^{(i+1)}-n_r^{(i)})=M \Delta V,
\end{equation}
is proportional to the potential difference $\Delta V$, where the
constant of proportionality $M$ corresponds to the number of sites 
per unit cell.

Thus, according to equation (\ref{ec_difu}), we can conclude that, given a periodic substrate
with unit cell of linear dimension $l$ and $M$ sites, the diffusion
coefficient and the potential difference between two  equivalent sites
in NN cells, are connected through the relation

\begin{equation}
  D = -j \frac{l^2}{M\Delta V},
    \label{D_V}
\end{equation}
where $j$ is the steady-state current density.

\section{Self-similar substrates}\label{sec:fractal}

Deterministic fractals are usually built by a recursive procedure, that
results in a sequence of structures called {\it generations}. 
A generation consists of a periodic array of sites connected by bonds. 
The process begins with a basic periodic structure (zeroth generation). At every step
the unit cell is scaled by a factor $L$ and the building rules ensure that
 self-similarity is obtained after a large number of iterations.

Following equation (\ref{D_V}), the diffusion coefficient $D_{p}$ for the
generation $p$ and the potential difference $\Delta V_{p}$ between two equivalent points in
 NN unit cells  are related as

\begin{equation}
    D_{p} = -j \frac{L^{2p}}{M_{p}\Delta V_{p}},
    \label{D_V_p}
\end{equation}
where $M_{p}$ is the number of sites in the unit cell, and  $L^{p}$ is its  linear
dimension. Then, for two consecutive generations $p$ and $p+1$, through which the same steady-state current flows,  we obtain

\begin{equation}
    \frac{D_{p}}{D_{p+1}}=  L^{-2}\frac{M_{p+1}}{M_{p}}
    \frac{\Delta V_{p+1}}{\Delta V_{p}}.
    \label{D_cociente_1}
\end{equation}

Now, since for a fractal the number of sites in a box with linear dimension 
$l$ behaves as $m(l)\sim l^{d_f}$ (i.~e., $d_f$ is the fractal dimension 
defined through box-counting), $M_{p+1}/M_{p}=(L^{(p+1)}/L^{p})^{d_f}=L^{d_f}$, 
and the last equation  can be rewritten as 

\begin{equation}
    \frac{D_{p}}{D_{p+1}}=  L^{d_f-2}
    \frac{\Delta V_{p+1}}{\Delta V_{p}},
    \label{D_cociente_2}
\end{equation}

As previously shown \cite{Padilla2009,Padilla2010}, a
perfect diffusive self-similar structure corresponds to a ratio
$D_{p}/D_{p+1}$ which does not depend on $p$, i.~e.,    

\begin{equation}
\frac{D_p}{D_{p+1}}=1+\lambda,
  \label{lambda}
\end{equation}
with $\lambda$ a positive constant. In this model, 
the mean-square displacement 
for a single random walker behaves as 
\begin{equation}
\Delta^2 x(t)=f(t) t^{2\nu}.
    \label{msd}
\end{equation}
The modulation $f(t)$ is a log-periodic function, $f(t \tau)=
f(t)$, and both $\nu$ and $\tau$ can be  analytically
calculated in terms of $L$ and $\lambda$:

\begin{equation}
    \nu =\frac{\displaystyle 1}{\displaystyle 2+\frac{\displaystyle
    \log(1+\lambda)}{\displaystyle \log(L)}}
    \label{nu_g}
\end{equation}
\begin{equation}
      \tau = L^{1/\nu}
      \label{tau_l}
\end{equation}

The important partial conclusion in the context of this work
is that, 
according to above discussion, a perfect diffusive
self-similar structure implies  a power-law behaviour 
for the resistance as a function of length. Indeed, equations (\ref{D_cociente_2}) and (\ref{lambda}) leads to

\begin{equation}
  \frac{\Delta V_{p+1}}{\Delta V_{p}}=L^{1/\nu-d_f},
  \label{DeltaV}
\end{equation}
where we have used $1+\lambda=L^{1/\nu-2}$, from equation (\ref{nu_g}).
Thus, for a perfect diffusive self-similar fractal the
potential difference, which corresponds to steady-state
current, scales with length $l$ as

\begin{equation}
  \Delta V \sim l^\zeta,
  \label{Delta_V_l}
\end{equation}
where the exponent $\zeta$ is given by

\begin{equation}
    \zeta=1/\nu-d_f;
    \label{Eins_perfect}
\end{equation}
which is the Einstein relation (\ref{Eins_1}), with $d_w=1/\nu$.

\section{Local exponents}\label{sec:local}

We consider now a generic substrate for which diffusive self-similarity is reached
only asymptotically. Let us assume a ratio between
consecutive diffusion coefficients, that
depends on the generation $p$,  as
\begin{equation}
\frac{D_{p}}{D_{p+1}}=1+\lambda_p.
  \label{lambda_loc}
\end{equation}
where, $\{\lambda_p:\, p=1,2,...\}$ is a sequence of
non-negative real numbers, with  $\displaystyle \lim_{p\to\infty}\lambda_p =
\lambda$. 

Because of this limit, at long enough times a single random
walk on this substrate will show a MSD behaviour as in
equation (\ref{msd}), and, as pointed out before,
 for large enough lengths  the potential
difference will behave as in equation (\ref{Delta_V_l}); with $\nu$ and $\zeta$ given
by equations (\ref{nu_g}) and (\ref{Eins_perfect}).

In this section we focus on  local exponents, which
correspond to the slopes in log-log scales for finite
length or time. As shown for example in \cite{Padilla2010}, on a substrate on which
diffusion coefficients for generations $p$ and $p+1$ satisfy
equation 
(\ref{lambda_loc}), the  MSD for a single random walker behaves as

\begin{equation}
    \Delta^2 x(t)\sim t^{2\nu_p},\;\;\;\mbox{for}\;\;\; L^p \lesssim \Delta x
    \lesssim L^{p+1}, 
    \label{msd-loc}
\end{equation}
with the local  exponent $\nu_p$ given by
\begin{equation}
    \nu_p =\frac{\displaystyle 1}{\displaystyle 2+\frac{\displaystyle
    \log(1+\lambda_p)}{\displaystyle \log(L)}}\;\;\cdot
    \label{nu_loc}
\end{equation}

Then, after rearranging this equation as $1+\lambda_p=L^{1/\nu_p-2}$, which 
corresponds to the  left hand side of equation (\ref{D_cociente_2}), 
we obtain

\begin{equation}
  \frac{\Delta V_{p+1}}{\Delta V_{p}}=L^{1/\nu_p-d_f}.
  \label{DeltaV_loc}
\end{equation}
Thus, we expect that the potential difference  scales with 
length $l$ as

\begin{equation}
    \Delta V(l) \sim l^{\zeta_p},\;\;\;\mbox{for}\;\;\; L^p \lesssim l
    \lesssim L^{p+1}, 
    \label{DV-loc}
\end{equation}
and that the local exponents satisfy the relation
\begin{equation}
  \zeta_p=1/\nu_p-d_f.
  \label{Eins_loc}
\end{equation}

Therefore, local slopes in log-log scales for the resistance as
a function of length and for  MSD of a single random walker
as a function of time are related for all scales through equation
(\ref{Eins_loc}); which generalizes the Einstein relation.

\section{Numerical simulations}\label{sec:numeric}
We study numerically the steady-state that corresponds to a unitary current on
two models, for which diffusive self-similarity appears asymptotically. At
finite lengths, the local random-walk exponent $\nu_p$ is not constant. 
Thus, we expect an also variable resistance exponent
$\zeta_p$, related to the former through equation (\ref{Eins_loc}). 

\begin{figure}[!ht]
	\begin{center}
    \includegraphics[scale=.4, clip]{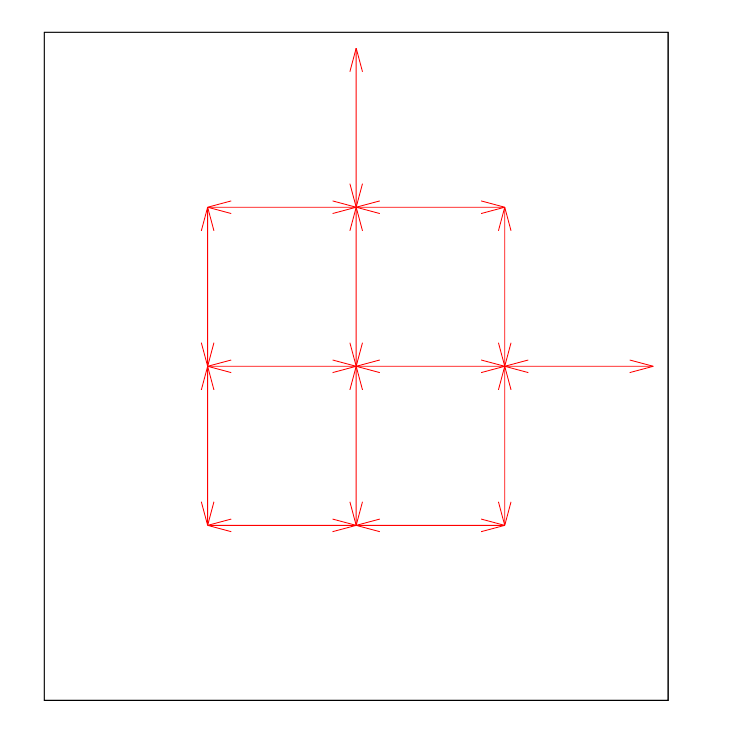}
		\hspace*{1.cm}	
	 \includegraphics[scale=.4,clip]{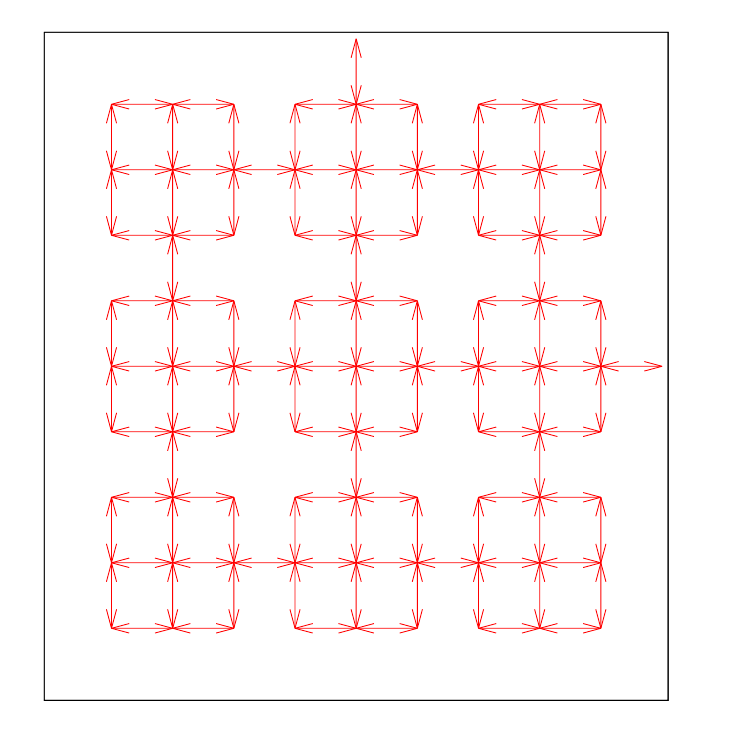}\\
 \includegraphics[scale=1.,clip]{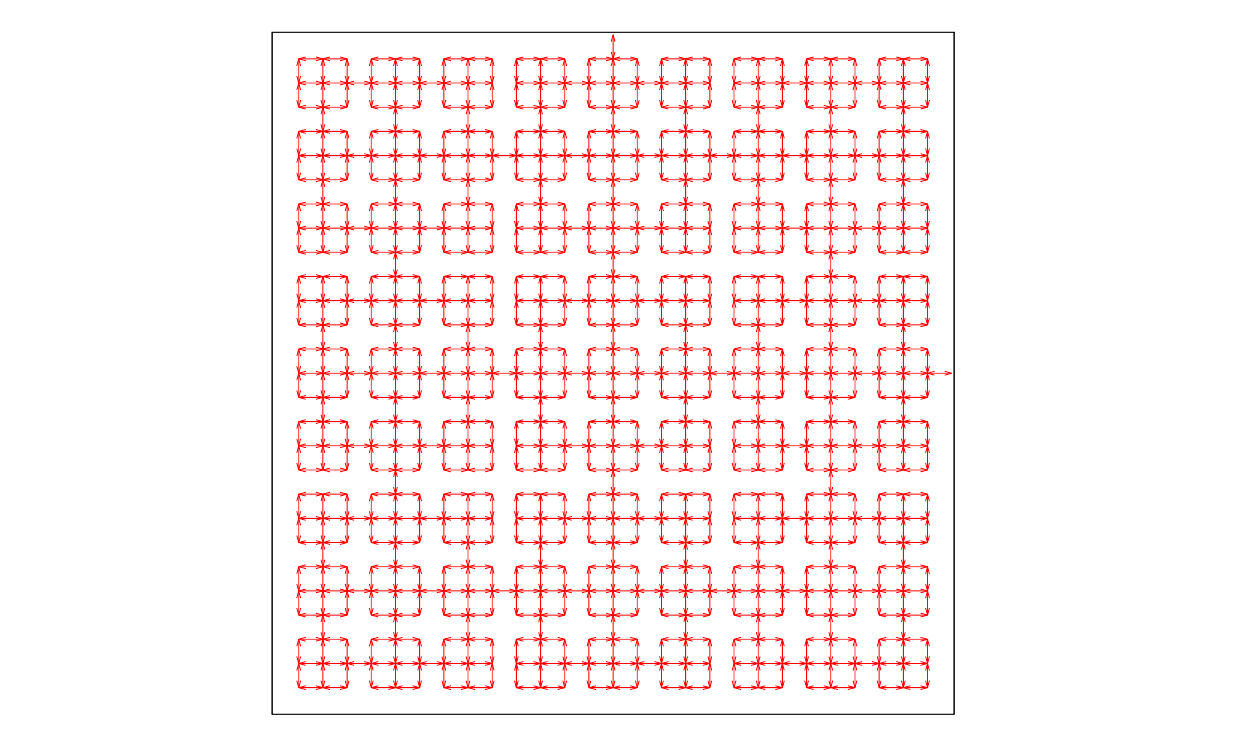}
 \end{center}
	\caption{Substrate in two dimensions,
    which results in scale-dependent  walk and resistance exponents.
    The schematics correspond to the unit cells for the first, second and third
    generations. The segments represent bonds between sites.}
\label{sustratos}
\end{figure}

The first model is a substrate built on a square lattice. A random walk 
consists in a particle hopping among NN sites.
If sites are connected by a bond, the hopping rate is $k=1/4$. If
the sites are not connected, the hopping rate is $k=0$. 
A fractal is obtained by deleting some bonds. The characteristic scale 
factor is $L=3$, and the unit cells for the
first, the second and the third generations are depicted
schematically in figure \ref{sustratos}. For every generation the unit cell can be separated from the
rest by cutting four bonds.  As shown in a previous work, the mass on this
structure shows a power-law  behaviour with $d_f=2$. However, the random walk
exponent $\nu_p$ grows with time  and approaches a value
$\nu<1/2$ when $t\to\infty$ \cite{Padilla2010}.

\begin{figure}[!ht]
	\begin{center}
	 \includegraphics[scale=.8,clip]{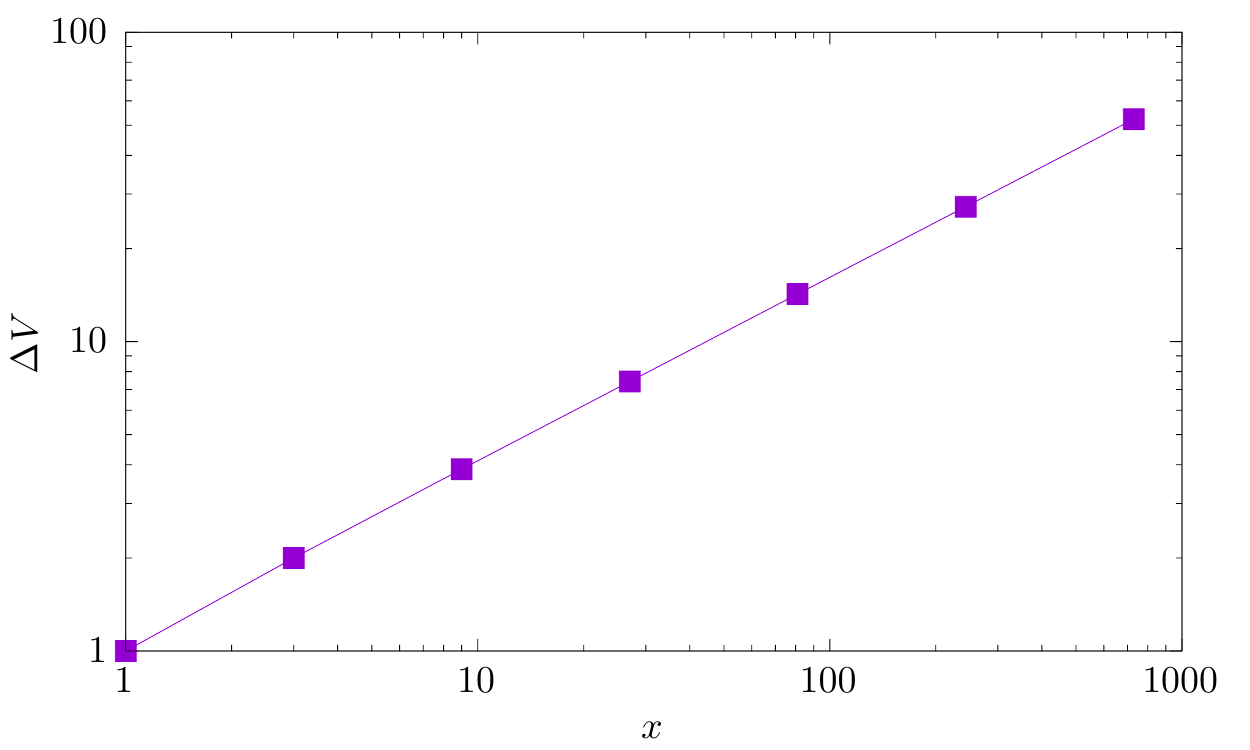}	  
 \end{center}
	\caption{Potential difference as a function of length for a unitary
     current flowing trough the unit cell of the sixth generation substrate in
     figure~\ref{sustratos}. The symbols correspond to simulations of the
     steady-state. The line was plotted with the exponents $\zeta_p$ from 
     equation (\ref{Eins_loc}) and the values of $\nu_p$ which result  from
     random-walk numerical simulations. 
	}
        \label{V_x_cua}
\end{figure}

We have run numerical simulations on the unit cell of the sixth generation,
to reach the steady-state in which a unitary current  flows between the 
left and right extremes. In figure~\ref{V_x_cua} we plot with symbols
the potential differences for lengths $x=3^i$ ($i=0,1,...,6$), which
are the unit cell linear sizes for the generations zero to
six. In the same figure, we plot a line using the relation (\ref{Eins_loc})
 and the numerical values for $\nu_p$, which are the
 outcomes of random walk simulations reported in reference~\cite{Padilla2010}. 
 Notice that both data set
 fall on  the same curve, which confirms the relation
 (\ref{Eins_loc}).

\begin{figure}[!ht]
	\begin{center}
	 \includegraphics[scale=.8,clip]{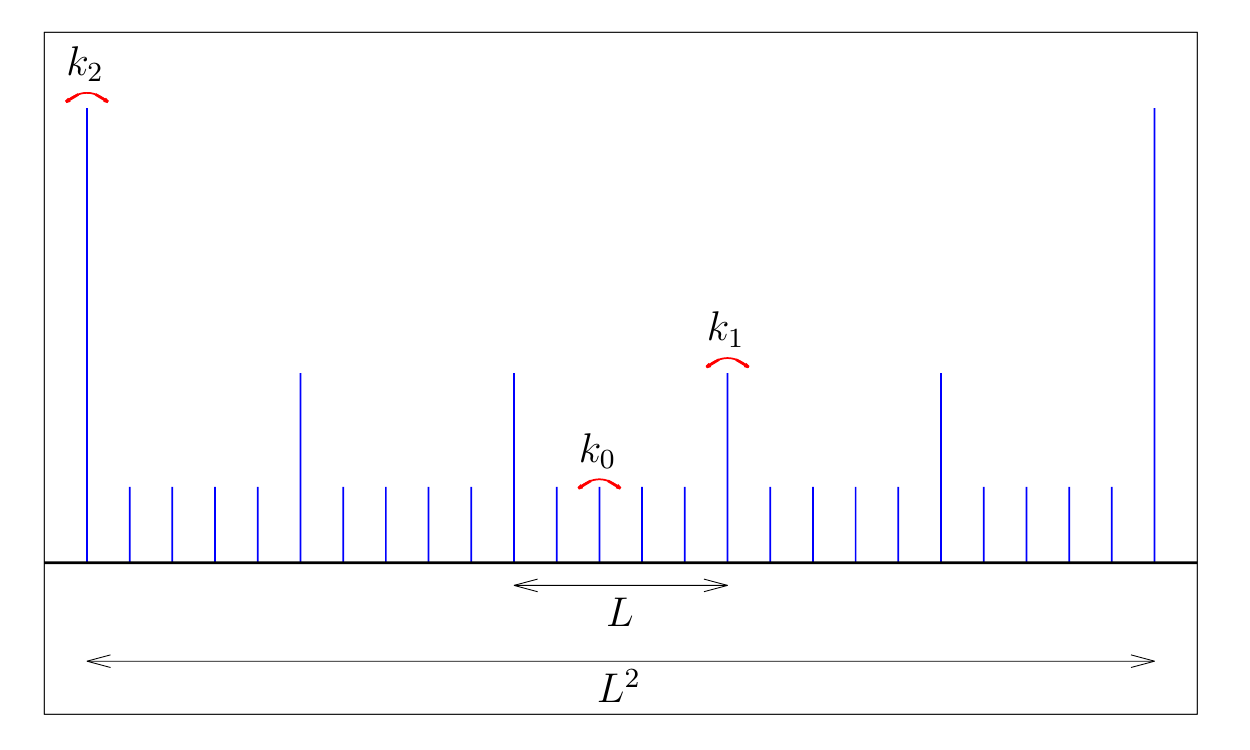}	  
 \end{center}
	\caption{Schematics of the one-dimensional random-walk model. 
    We begin with a homogeneous lattice, and a hopping rate $k_0$ between 
    nearest-neighbor sites. Then, hopping rates are reset to $k_j$ for 
    transitions between sites $j$ and $j+1$ for every $j$ such that
    $mod(j,L^n)=0$,  and for $n=1,2,...$. In this example, $L=5$.
	}
    \label{sustrato_1D}
\end{figure}

The second model is a generalization of the one-dimensional self-similar 
model introduced in \cite{Padilla2009}. We start with a single random walk
on a one-dimensional lattice, with a hopping rate $k_0$ between any pair of
NN sites. This homogeneous case corresponds to generation 
zero. We introduce a natural number $L$ to build the other
generations. 

In the first generation, we reset to $k_1<k_0$ the hopping rate 
 for every pair of sites $j$ and $j+1$, with 
 $mod(j,L)=0$. The other hopping rates
 remains as in zeroth generation. 

 In the second generation, we reset to $k_2<k_1$ the hopping rate 
 for every pair of sites $j$ and $j+1$, with 
 $mod(j,L^2)=0$. The other hopping rates
 remains as in first generation. 

This recursion follows indefinitely, in such a way that generation
$n$ is obtained from generation $n-1$ after resetting to $k_n<k_{n-1}$
the hopping rate 
 for every pair of sites $j$ and $j+1$, with 
 $mod(j,L^n)=0$. In figure~\ref{sustrato_1D} we show an schematics for $L=5$.

If we ask for perfect self-similarity for diffusion, i.~e. equation
(\ref{lambda}), the hopping rates are found iteratively as in reference 
\cite{Padilla2009}. For the more general case of equation (\ref{lambda_loc}),
the sequence of hopping rates is given by

\begin{equation}
  \frac{1}{k_{i}}=\frac{1}{k_{i-1}}+\frac{L^i\lambda_{i-1}}{k_0}
  \prod_{j=0}^{i-2}(1+\lambda_j),\;\;\;\;\;\mbox{for}\;\; i=1,2,3...
  \label{k_i}
\end{equation}

We test the validity of the relation (\ref{Eins_loc})
among the local exponents for a  family of substrates
 given by 
\begin{equation}
  \lambda_p=\lambda\,  (1-2^{-p/5.}).
  \label{lambda_p_1d}
\end{equation}
At short enough lengths  these substrates are nearly homogeneous
($\lambda_p\approx 0$ for $p\ll 5$), while,  on the other extreme,
 self-similarity for diffusion is reached for lengths 
much larger than $L^5$. The local random walk 
exponent (\ref{nu_loc})  decreases with length and approaches
asymptotically $\nu$ in equation (\ref{nu_g}). Thus,   the 
variation of $\nu_p$ in space increases with $\lambda$ and, because of 
equation (\ref{Eins_loc}), the same should occur with the variation of
$\zeta_p$. This is an interesting model, because the variation of the
exponents with length can be adjusted  through the parameter
$\lambda$.

\begin{figure}[!ht]
	\begin{center}
	 \includegraphics[scale=.8,clip]{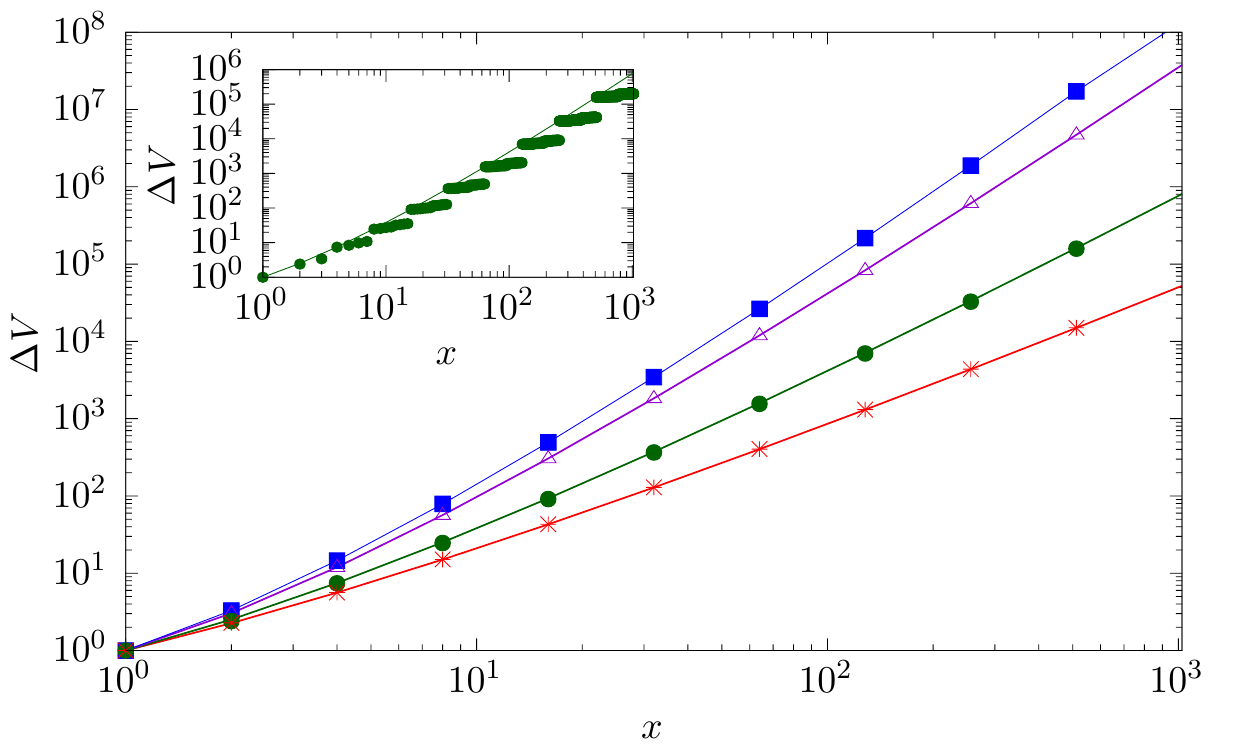}	  
 \end{center}
	\caption{Potential difference as a function of length for unitary 
    current on the one-dimensional model with $\lambda_p=\lambda\,
    (1-2^{-p/5.})$, and $L=2$. 
    (Main) Symbols correspond to data obtained with numerical simulations on a tenth-generation
    substrate. Lines were drawn using the values of theoretical exponents.
    From bottom to top, $\lambda = 1$ (red), $\lambda = 2$ (green), $\lambda
    =4$ (violet), $\lambda=5$ (blue).  
    (Inset) More detailed structure for $\lambda=2$.
	}
    \label{V_x_1D}
\end{figure}

We have run numerical simulations for the steady-state that corresponds to a
unitary current flowing on this model, with
$L=2$ and $\lambda=1, 2, 4, 5$. All substrates were built until generation $10$.
In figure~\ref{V_x_1D}-main we plot with symbols the potential difference as
a function of the length $x$, for  $x=2^j$ ($j=0, 1,...,9$).
The lines correspond to the exponents $\zeta_{p}$ obtained from equations
(\ref{Eins_loc}) and (\ref{nu_loc}). Note the excellent  agreement between
theory and simulations. The inset in the same figure shows 
substructure of $\Delta V$ for  $\lambda=2$.

\section{Conclusions}\label{sec:conclu}
We have studied first the connection between  single random walks and
steady-state  potential difference for substrates with spatial periodicity.
Then, by considering a sequence of periodic systems, a common procedure  
for deterministic fractal construction, we find  that the length 
dependent fractal,
walk and resistance exponents, for the substrate obtained in the infinite
limit of this sequence, satisfy, at every length scale, the relation
(\ref{Eins_loc}). This  can be considered as a local version of the 
 Einstein relation (\ref{Eins_1}).
 We have tested our predictions numerically for two models. The first
 model is a fractal 
 in two dimensions, while the the second is a fractal in one dimension.
 Both models lead to 
 length-dependent exponents at intermediate scales. The excellent agreement between 
 the outcomes of these simulations and the theoretical predictions
 supports the validity of the mentioned relation among  exponents,
 not only in the asymptotic self-similar limit but also locally, for
 all length scales.

\section*{Acknowledgments}
We are grateful to H.~O.~M\'artin for useful discussions.
This research was supported by the Universidad Nacional de Mar del Plata,
15/E1040, and the Consejo Nacional de Investigaciones
Cient\'{\i}ficas y  T\'ecnicas, PIP1748/21.
\section*{References}



\end{document}